Davide Neri[1]

Scuola di Dottorato in Antropologia ed Epistemologia della Complessità
Centro studi storici transdisciplinari sulle scienze e la filosofia - ISHTAR
Università di Bergamo – Italy

# Between Being and Becoming:
# a meta-theoretical approach to irreversibility

**Abstract** – The relationship between reversible-dynamical and irreversible-thermodynamic descriptions is analyzed from a meta-theoretical point of view. A network of inter-theoretical relations is drawn by means of asymptotic relations and complementarity. It is argued that the "paradox" between micro-reversibility and macro-irreversibility can be overcome by assuming a meta-complementarity relation existing between different levels of physical description, preserving a realist conception.

**Keywords** – Irreversibility, complementarity, inter-theory relations

---

[1] Present address: Liceo Scientifico "A.B. Sabin", via Matteotti, 7 – 40129 – Bologna. E-mail address: davide.neri@istruzione.it

## 1. Introduction

The contrast between the reversibility of the laws of dynamics and the thermodynamic irreversibility is still at the center of a debate that shows no sign of closing. On the one hand there is a long tradition, started with the definition of entropy itself, that attempts to deduce the macroscopic irreversibility as a result of the statistical behavior of a large number of elementary constituents which obey to the laws of reversible dynamical, classical or quantum. This tradition has received its first contribution by Maxwell and Boltzmann.

On the other hand, there is the proposal to modify the dynamical description, in order to introduce into the microscopic level an irreversible behavior that should be reproduced on a large scale. The Prigogine's work may be brought back to this research program.

However, both approaches, that are essentially based on the reduction of the macroscopic to the microscopic level, encounter some difficulties. The former, despite its considerable success in the description of many macroscopic phenomena, is unable to obtain a full deduction of the irreversibility from the reversible dynamics and must to confront with the emergence of collective properties not reducible to the fundamental level; the second presents various elements of arbitrariness and must have a reckoning with the fact that the thermodynamic level seems to be independent on the reversible or irreversible behavior of single atoms and molecules.

This paper aims at overcome the problem through a meta-theoretical analysis that leads to recognize the existence of multiple levels of description of physical reality, each of them having its own autonomy, and complementary each other. The solution proposed here would preserve a realist conception of nature, based on the following points:

- The belief that the world has been created for us belongs to a tradition that does not find many confirmations in the results of contemporary science. In the past centuries, it led to assume the existence of a pre-established harmony between the human mind and physical reality and, consequently, to invoke the possibility (or the necessity) of a unique and complete description of the world. This presupposed harmony has played an important role in the development of the reductionist research programs, that are very far from reaching their intended purpose.

- On the contrary, all the conceptual structures that we use to order the phenomena exhibit a historical/evolutionary character and are continuously adapted to the limitations imposed by the outside world;
- As we are immersed into the physical world and we are a part of it, we can only obtain partial views of reality: it is impossible, for us, to have the *God's eye perspective* of the world that a long tradition of the scientific thought has borrowed from the religious attitude;
- Then, it becomes crucial to investigate which relations exist between these views (partially true and partially contradictory) of reality;
- The pattern of complementarity, which will be employed here, and which has been repeatedly accused to be incompatible with a realist conception, may be used to argue in favor of a "well-tempered realism";
- In this framework, the inter-theoretical analysis becomes fundamental for the understanding of the multiplicity of levels of description requested by a complex physical reality.

## 2. The mechanistic hypothesis: Boltzmann et al.

The relation between dynamical reversibility and thermodynamic irreversibility has been the topic of a long-time debate since the introduction of the entropy concept by Rudolf Clausius in 1865. Many attempts have been made to explain the macroscopic irreversibility as a consequence of the behavior of the very large number of atoms and molecules contained in macroscopic bodies.

In XIXth century these attempts have been pursued according to different approaches that tried to obtain:

a] a direct or an analogical deduction of the second law of thermodynamics from the Hamilton equations of motion (see for example Boltzmann 1866, Helmholtz 1884);

b] a statistical interpretation of the second law of thermodynamics based on the Boltzmann's law $S = k \ln W$ (Boltzmann 1872), that introduces a statistical generalization of thermodynamic entropy for a single system.[2]

---

[2] In this paper I will not deal with the theory of statistical ensembles because, notwithstanding its successes in dealing with macroscopic equilibrium and the many thermodynamic analogies that it allows to obtain, it can treat the irreversible phenomena only thanks to the introduction of *ad hoc* hypotheses

Within the approach a], Boltzmann (1866) tried to deduce the second law as a direct consequence of dynamical equations of motion, but he could only obtain a result limited to a very particular system: a gas that performed a periodical motion with period τ. In this case, that corresponds to a totally degenerate multiperiodic mechanical system (Born 1969, Appendix XIII), he obtained a mechanical quantity analogous to the entropy, through an equation having the form $S = k \ln J$, where $J$ is the sole adiabatic invariant of the gas. The equation reveals an interesting link between entropy and action and between their adiabatic invariance properties. However, any attempt to generalize Boltzmann's result has to deal with the validity of more or less restrictive hypotheses of ergodicity (see **4.2**).

The approach b] led to the development of Statistical Mechanics that currently makes understandable many thermal phenomena, but it is unable to give a complete reduction of thermodynamics to mechanics. In the Boltzmann approach the proof of the H-theorem depends on the *Stoßzahlansatz*, a hypothesis which is not verified by *all* the dynamical states of the system and could become valid for *almost all* the initial conditions only in the Boltzmann-Grad limit.

Instead of the term Statistical Mechanics, in the following, I prefer to use the term Statistical Thermodynamics (ST) to indicate the whole of branches that have been developed to generalize the pure Classical Thermodynamics (CT) in agreement with the discovery of the atomic structure of matter. The transition from the pure and phenomenological thermodynamic level to the statistical one is characterized by the identification of the gas constant $R$ with the product $N_A k$, where $N_A$ is the Avogadro number and $k$ is the Boltzmann constant.

The statistical interpretation of the second law (Boltzmann 1872 and 1877) cannot give us a mechanical deduction of CT. The irreversible behavior of macroscopic systems is explained as the most probable behavior of a mechanical system, but it has an intrinsic probabilistic character that cannot exclude the existence of violations of the second law. As Maxwell (1870) said,

> "The 2$^{nd}$ law of thermodynamics has the same degree of truth as the statement that if you throw a tumblerful of water into the sea, you cannot get the same tumblerful of water out again."

---

(such as the coarse-graining).

This circumstance gives rise to the problem of the relationship between the dynamical description and the use of probability. According to a long tradition, which recognizes a character of reality only to the microscopic level, the emergent properties of macroscopic bodies are considered to be apparent. The irreversibility, for example, should only occur as a result of our incomplete knowledge of the state of dynamic systems and would therefore be the direct result of our ignorance, because only exist one level of reality can exist and all the phenomena that conflict with it must be apparent. Perhaps the most explicit statement of this attitude was given by Born, who defined it as the "method of ignorance" (Born 1949, p. 68).

Anyhow, neither approach a] nor approach b] lead to a full deduction of thermodynamic laws from the dynamical equations of motion. They can only establish formal analogies and/or asymptotic correspondence rules between the macroscopic and microscopic behavior and reveal that the reduction of CT to CM is a very hard problem. The situation has been described in a very effective way with the words written by Poincaré in 1893:

> "The problem [of the deduction of macro-irreversibility from micro-reversibility] is so complicated that it is impossible to treat it with complete rigor […], but there is no need for a long discussion in order to challenge an argument of which the premises are apparently in contradiction with the conclusion, where one finds in effect reversibility in the premises and irreversibility in the conclusion." (Poincaré 1893, p. 537)

**3. From Being to Becoming: Prigogine**

If the reduction of macroscopic irreversibility to microscopic reversible laws is impossible, one might be tempted to modify the dynamical laws in order to have the macroscopic irreversibility as a logical consequence of a (supposed) microscopic irreversibility. This choice can be encouraged by the need to avoid any reference to anthropomorphic elements in order to explain the emergence of the irreversibility at the macroscopic level, as that employed by the method of ignorance.

Prigogine and its school have gone along this way. Prigogine tried to find in the microscopic level the roots of irreversibility by postulating the substitution of the fundamental concepts of classical dynamical description (point-phases and trajectories) with the recourse to distribution functions and statistical ensembles. The rejection of point-phases and trajectories has been motivated by their too abstract nature.[3]

However, Prigogine himself must admit that, according his generalization,

> "there is still a trajectory if initial conditions are known with infinite precision. But this does not correspond to any realistic situation. When we perform an experiment, whether by computer or some other means, we are dealing with situations in which the initial conditions are given with a finite precision and lead, for chaotic systems, to a breaking of time symmetry." (Prigogine 1997, p.105)

The substitution is originated by practical considerations and do not imply changes in the conception of physical reality. But it is tantamount to admitting that the introduction of the distribution functions is motivated by considerations of epistemic and not ontological kind: they represent a convenient solution that serves to overcome practical difficulties, without affecting the nature of things. From this point of view, the Prigogine's position does not represent a true alternative to the "method of ignorance".

If this wanted to avoid the introduction of subjective interpretations of thermodynamics and irreversibility, the result would be to transfer this subjectivity to the microscopic level, without scratching the traditional reductionism that tends to consider the macroscopic as a secondary effect of the microscopic level. The Prigogine's attitude is not less reductionist than that of Born, even if some of his texts reveal a lot of ambiguity:

> "Irreversibility is either true on all levels or on none. It cannot emerge as if by a miracle, by going from one level to another." (Prigogine and Stengers 1985, p. 285)

> "The existence of phase transitions shows that we have to be careful when we adopt a reductionist attitude. Phase transitions corresponds to emerging properties. They are meaningful only at the level of populations, and not of single particles. It is at the level of populations that the symmetry between past and future is broken, and science can recognize the flow of time" (Prigogine 1997, p. 45)

---

[3] As for the abstraction, the use of statistical ensembles that characterize the theory of Gibbs should be handled with caution: in fact, we never observe statistical ensembles, but only individual macroscopic systems that follow the Boltzmann's law. Therefore, these ensembles are not more concrete physical objects than the trajectories of the classical dynamics.

On the one hand Prigogine admits that the phase transitions are an emerging phenomenon, to which we can make sense only at the macroscopic level and in close relationship with the thermodynamic limit, but he refuses to assign the same character to macroscopic irreversibility, believing that its origin is to be sought at the microscopic level, in a modification/generalization of the current dynamic description. He then moves in a reductionist perspective but, as well as the traditional reductionist-mechanist which make use of coarse-graining and the method of ignorance (as Born), he is forced to adopt an ambiguous language and to confuse the ontological and the epistemic levels.

## 4. A third way: "between Being and Becoming"

Recent works (Dettman e Cohen 2000 and 2001, Cecconi & al. 2003) on the behavior of (chaotic) Lorentz gas and (not-chaotic) Ehrenfest wind-tree model confirm that the macroscopic irreversible phenomena, as diffusion, occur independently on the dynamic behavior of microscopic constituents of matter. It could be objected that numerical simulations made with computer imply the use of rational numbers, and this could introduce an element of irreversibility in any kind of model, independently on its chaotic properties, but it is well known from its introduction that the Ehrenfest wind tree model (Ehrenfest and Ehrenfest, 1912, Appendix to Sect. 5) has an irreversible evolution for almost all the initial conditions, in full agreement with the Boltzmann statistical approach.

The macroscopic irreversibility appears to be insensitive to the microscopic laws of motion and we must recognize the autonomy of the macroscopic level. As well as Prigogine, I think that the macroscopic phenomena are real, at least as the microscopic ones, but I claim too that the "paradox" of irreversibility cannot be resolved by appealing to some kind of reductionism (from *micro* to *macro* or vice versa).

### 4.1 A meta-theoretical scheme

The solution that is proposed here to overcome the problem of the contrast between reversibility and irreversibility is based on the existence of the network of inter-theoretical relations presented in the following scheme (Figure 1):

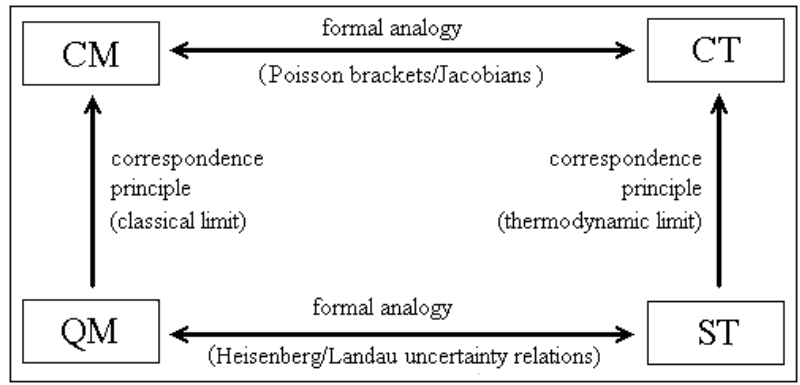

**Figure 1** – Formal analogies and correspondence principles

It relates four theories – classical mechanics (CM), quantum mechanics (QM), classical thermodynamics (CT) and statistical thermodynamics (ST) – and contains four binary relations that need some explanation.

Obviously, it would be very superficial and imprudent to say that this scheme runs out the relationships among the theories involved. They have complex structures that cannot be superimposed one on the others, and their inter-theoretic relations are, to the same extent, very complex. The scheme, that includes inter-theoretic reductions and formal analogies, can appear as an oversimplification, especially because the ST-CT and QM-CM relations are a lot more intricate than is commonly thought (Sklar 1993, Bokulich 2008).

However, among the theories it is possible to find some common elements that throw some light on their relations. In the following pages we will have to assess whether the positive analogies are sufficiently strong.

1] *The relation CM-CT*. It is well-known that these theories are based on the canonical formalism (Corber and Stehle 1950, Peterson 1979, Rey de Luna and Zamora 1986): in both theories we can find pairs of conjugated quantities, that are linked by Poisson brackets in CM and by jacobians in CT. For mechanical conjugated variables we have $[q,p]=1$ and in CT, for conjugated intensive and extensive quantities, we obtain $J(E,1/T)=1$ and $J(V,P/T)=1$ (in the entropy representation). The equations of motions have a thermodynamic analogous in the Maxwell relations and a correspondence between action and entropy may be deduced on the basis of their properties of adiabatic invariance (Leff 1995, Bellan 2004). See the following **4.2**.

2] *The relation QM-CM*. The classical limit may be expressed by the following condition imposed on the Planck constant $h$ and the quantum number $n$:

$$h \to 0, n \to \infty, \text{ with } nh = \text{constant}.$$

The necessity of both conditions on $h$ and $n$ has been discussed in many relevant physical problems (Hassoun and Kobe 1989). In this limit, the Heisenberg relations become negligible: $\Delta q \Delta p \to 0$. According to the complementarity relation introduced by Bohr (1928) between causal and space-time descriptions, this limit corresponds to the fall of the limitations imposed on the definition of dynamical concepts by the existence of the quantum of action. The condition $h \to 0$ removes the quantum uncertainties on the exact definition of a classical dynamical state, given by the exact values of coordinates and momenta. It must be stressed that, according to Bohr, the quantum limitations are the consequence of the introduction of the Planck constant, which he considered an inescapable element of reality.

3] *The relation QM-ST*. Into the framework of equilibrium ST it is possible to deduce, from the fluctuation theory of macroscopic quantities of a system interacting with a large (but finite) reservoir, some uncertainty relations having the same form of Heisenberg uncertainty relations of QM. In the entropy representation we have:

$$\Delta E \Delta (1/T) \sim k \text{ and } \Delta V \Delta (P/T) \sim k.$$

In this case, the Boltzmann constant plays the same role of the Planck constant in QM. These relations, that I will call Landau uncertainty relations (Landau & Lifschitz 1958, chapter XII), have been analyzed by many authors, but their interpretation is still object of a long debate (see for example Uffink and van Lith 1999). From an analogical point of view, their meaning can be related with an uncertainty inherent the definition of the macroscopic equilibrium state due to the fluctuations induced by the statistical behavior of atoms and molecules at the microscopic level. This point will be discussed in **4.3**.

4] *The relation ST-CT*. Usually the relation between these theories is based on the definition of a thermodynamic limit expressed in the form: $N \to \infty$, $V \to \infty$, $E \to \infty$, with $V/N$=constant and $E/N$=constant. These conditions act on the extensive properties of the systems leaving unchanged the intensive ones. An alternative definition can be given by the conditions:

$$k \to 0, N_A \to \infty, \text{ with } N_A k = \text{constant and } N_A m = \text{constant},$$

where $N_A$ is the Avogardo number and $m$ is the mass of a single molecule (Neri 1987

and 1993). It leaves unchanged all the intensive and extensive variables of the thermodynamic system and acts on the microscopic structure. According both definitions, in thermodynamic limit the ideal gas law is exactly satisfied, and their formal equivalence has been investigated in some relevant cases (Compagner 1989), but the latter has the advantage to put in evidence an important analogy with the conditions expressed by the classical limit of QM. The condition $k \rightarrow 0$ imposed on the Landau uncertainty relations has the same effect of the condition $h \rightarrow 0$ on the Heisenberg uncertainty relations.

If at the limits $k \rightarrow 0$ and $N_A \rightarrow \infty$ we add the condition imposed by the Boltzmann-Grad limit on the radius $r$ of molecules ($Nr^2$=constant), the validity of the *Stoßzahlansatz* is preserved when $N$ diverges, and the statistical deviations from the "absolute" thermodynamic behavior become more and more negligible (Grad 1958).

## 4.2 Adiabatic analogy

The strict relation between adiabatic transformations in CM and in CT has been investigated by many authors. They have shown the existence of a direct correspondence between these processes and between the role played by action in CM and entropy in equilibrium CT, respectively.

Perhaps, the first important result on this topic can be ascribed to Boltzmann himself (1866). In its first paper on the mechanical interpretation of the second law of CT he tried to obtain a complete reduction on the Second Law of CT to Hamilton equations, but he only could find a mechanical analogous of entropy for a very particular system: a gas that performed a periodic motion in a time $\tau$. For that system, Boltzmann could write the following equation: $S = \ln T\tau$, where $T$ is the average of the kinetic energy of a single molecule calculated over the period $\tau$. It is easy to see that $T\tau$ is the unique adiabatic invariant of this system, because all the periods of the motions of single molecules are submultiples of the period $\tau$ (Born 1969, Appendix XIII).

A simple application of Boltzmann's result can be found for a one-dimensional gas composed by a single particle that bounces back and forth between two walls situated at a distance L. In this case the quantity $T\tau$ corresponds to the action integral $J=\oint pdq$ and has the physical dimensions of an action. This one-dimensional gas is a periodic and ergodic system, and *J* is its sole adiabatic invariant. A generalization of the Boltzmann relation $S = \ln J$ of a system composed by $N \gg 1$ molecules is given by the volume-entropy for the microcanonical ensemble, $S = k \ln \Phi$, if some additional conditions are verified. The quantity $\Phi$ has the same geometrical meaning of the action integral *J* for the one-dimensional gas: it represents the volume closed by the surface of energy *E* into the phase-space of the system.

Hertz (1910) proved that, if the system satisfies the ergodic hypothesis, $\Phi$ is an adiabatic invariant, as well as *J*. Campisi and Kobe (2010) have shown that, to ensure the adiabatic invariance of $\Phi$, it is sufficient the validity of two conditions of weak-ergodicity. The adiabatic analogy between action and entropy in the one-dimensional gas has been stressed also by Leff (1995) and Bellan (2004).

**4.3 Canonical formalism, uncertainty relations and complementarity**

As it has been said in **4.1**, both CM and equilibrium-CT have a mathematical structure that is based on the canonical formalism. These theories are founded on a dualistic structure, in which the state of the system depends on pairs of conjugated quantities (coordinates and momenta in CM, intensive and extensive quantities in CT) that are linked by commutation relations: the Poisson brackets in CM, the jacobians in CT. Obviously, we cannot prove if this dualism is the consequence of the conceptual schemes that we use to order physical phenomena, or a property of the physical world itself: our answer depends on our philosophical preferences about the structural empiricism and the structural realism. However, the analogy shows an important evolution when we take into account the transitions from CM to QM, and from CT to ST. According to these theories, the conjugated quantities are subjected to the Heisenberg and Landau uncertainty relations. So, it seems that the dualism typical of the canonical formalism meets with some limitation to the definition of "classical" concepts and descriptions. The appearance of these limitation is related the discovery of the physical constants *h* and *k*, that appear as natural constraints imposed on the applicability of the canonical formalism to physical phenomena. The canonical formalism, that expresses with good approximation a large part of physical phenomena, must be adapted to the existence of the constants *h* and *k*, that seem to indicate the presence of an inescapable element of reality.

The formal analogy between Heisenberg and Landau uncertainty relations poses the problem of the interpretation of the latter. Bohr (1932) and Rosenfeld (1961) have proposed an interpretation according to which Landau relations should be the proof of

- a complementarity existing between micro-reversibility and macro-irreversibility;
- a complementarity between the possibility to obtain a definition of the temperature (that requires the interaction of the system with its surroundings) and a definition of the energy (that presupposes the insulation of the system).

According to Rosenfeld's, these complementarities should be two different expressions of the same relations.

It is not easy to follow Bohr's and Rosenfeld's suggestion in its original form, because the two complementarities refer to two different level of discourse: the first one inherits to the micro-macro relation, the other is related only to the macroscopic level of description, and the Landau relations refer only to the latter. In fact, they connect the fluctuations of macroscopic quantities to the value of the Boltzmann constant, which takes into account the existence of a discrete structure of matter. They imply the loss of the absolute character of the laws of CT and lead to a generalized concept of equilibrium, which does not correspond to a single invariable state of the system, but to a set of states with high entropy where the system persist nearly all the time.

The necessity of a rational generalization of the concept of thermodynamic equilibrium from a statistical point of view has been already suggested by other authors from different points of view (van Lith 1999, Lavis 2005).

From our analogical perspective, the Landau relations seem to be the expression of an uncertainty in the definition of the macroscopic equilibrium state, as well as the Heisenberg relations can be considered the measure of an uncertainty in the definition of classical state from a semi-classical point of view. They can be related to a complementarity existing between macroscopic quantities: the extensive variables, that give the size of the macroscopic bodies, and the intensive ones, that define their equilibrium conditions from a thermodynamic point of view. The Boltzmann constant, which here plays the same role of the Planck constant in Heisenberg relations, both represents the uncertainty in the definition of these macroscopic variables due to the discrete structure of matter and gives a measure of the stability conditions of thermodynamic systems (Einstein 1904).

The meaning of this uncertainty is illustrated in Figure 2, that shows the evolution of the entropy of an Ehrenfest urn model when $N\to\infty$. This condition, for a system with a fixed size, is equivalent to the limits $N_A\to\infty$ (and $k\to 0$). The fluctuations near the maximum value of entropy become more and more negligible and the system shows a pure thermodynamic behavior. If $N$ is finite the system does not have a definite equilibrium state, but its entropy remains almost all the time in an "equilibrium band" near to the maximum value. A unique characterization of equilibrium requested by CT can be obtained only when $N$ becomes infinite.

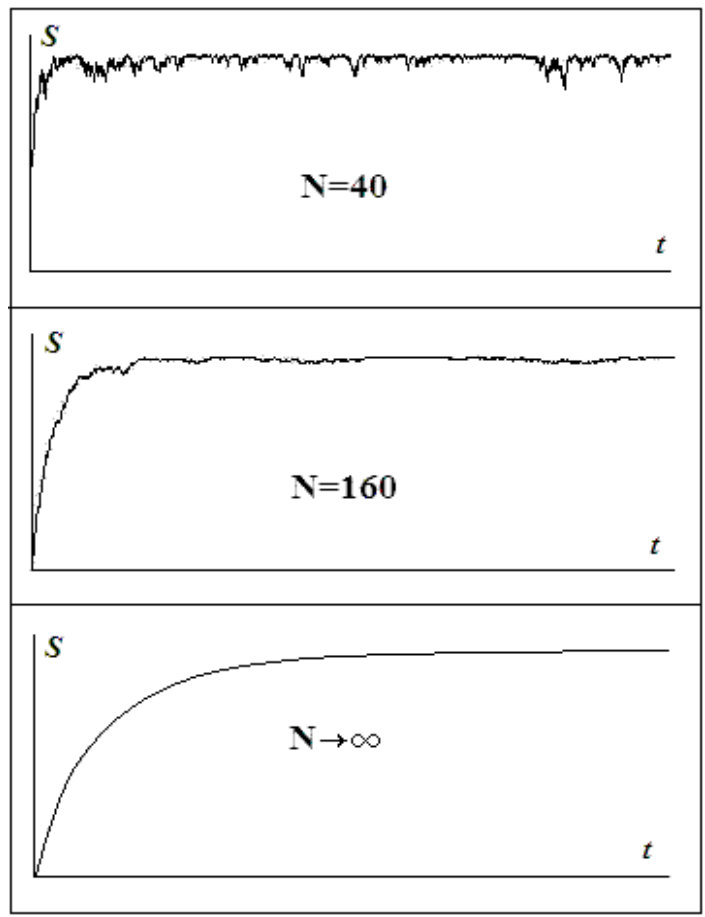

**Figure 2** – The increasing of entropy for Ehrenfest urn models with different $N$ values

There is an evident correspondence between the uncertainty in the equilibrium state and the statistical behavior implied by the introduction of the Boltzmann constant and the Avogadro number. Also, it is evident that, when $k \to 0$ and $N_A \to \infty$, the disappearance of the statistical behavior leads to a non-ambiguous definition of the equilibrium state.

## 5. Meta-complementarity and objective interpretation of probability in ST

At this point, it seems that we have a thermodynamic complementarity that acts at a macroscopic level, as well as the quantum complementarity is typical of the microscopic scale. However, contrary to the Bohr's first opinion, this thermodynamic complementarity cannot be used for the interpretation of the "micro-macro" relationship, because it pertains only to the macroscopic level. The micro-macro relationship needs another interpretation and another form of complementarity, that we can obtain only by means of a meta-theoretical analysis.

As we have seen, a coherent application of the "method of ignorance" has serious implications for our understanding of the physical world. It relegates the macroscopic phenomena to a level of contingency and pure appearance, in agreement with a reductionist vision, which claims

> "to explain the complication of the visible in terms of invisible simplicity" (Perrin 1916, p. VII).

Who believes in the reversible dynamical laws as the only true and objective laws of nature cannot accept the full objectivity of macroscopic phenomena and their emergent properties, as is clearly affirmed by Einstein in its well-known letter to Besso's widow:

> "People like us, who believe in physics, know that the distinction between past, present and future is only a stubbornly persistent illusion." (Einstein and Besso 1972, pp. 537-538).

However, the emergence of these phenomena can also be seen in another way, as the symptom of their essential irreducibility, and we need a very dogmatic line of thought to say that these emerging phenomena are only apparent. Rather, their irreducibility to the dynamical level could reveal their reality and the necessity to go beyond the reductionist conception. In this direction moves the Schrödinger's thought:

> "In text-books of gas-theory it has become a stock phrase, that statistical methods are imposed on us by our ignorance of the initial co-ordinates and velocities of the single atoms – and by the insurmountable intricacy of integrating $10^{23}$ differential

> equations, even if we knew the initial values.
>
> But inadvertently, as it were, the attitude changes. It dawns upon us that the individual case is entirely devoid of interest, whether detailed information about it is obtainable or not, whether the mathematical problem it sets can be coped with or not. We realize that even if it could be done, we should have to follow up thousands of individual cases and could eventually made no better use of than compound them into statistical enunciation. The working of the statistical mechanism itself is what we are really interested in." (Schrödinger 1944, p.704)

Karl Popper (1982) put in clear evidence the absurdities that we encounter if we adopt in ST the subjective interpretation of probability derived from the method of ignorance. The objectivity of thermal irreversible phenomena (for example, the diffusion of a gas from a bottle to a larger volume) cannot be confused with the subjective elements of our incomplete knowledge: the diffusion of a gas would occur also if we had a complete information about the states of all its molecules and, to explain it, we could not use the laws of dynamics, but we should make use of statistical laws. Moreover, if we had a limited knowledge of the dynamical state, our ignorance could only increase, but the same is not true for the entropy of a single system that, according to the Boltzmann's law, can fluctuate and decrease with negligible (but not *equal to zero*) probability. Ignorance of the dynamical state and entropy show a misleading analogy

The statistical description of macroscopic systems has an autonomous character and must be regarded as an alternative, complementary mode of description respect to the dynamical one. The recognition of this autonomy could be the keystone that removes the doubts expressed by Sklar:

> "the additional elements necessary to place thermodynamics into the general dynamic picture are clear, although […] much debate remains about how, exactly, they are to be construed. What is striking is that this additional posit seems to be itself so autonomous and so out of place in the remainder of our world picture." (Sklar 1993, p. 372)

## 6. Conclusion

The meta-theoretical network analyzed in chapter 3 may now be interpreted by means of two different levels of complementarity, that are shown in the following scheme (Figure 3):

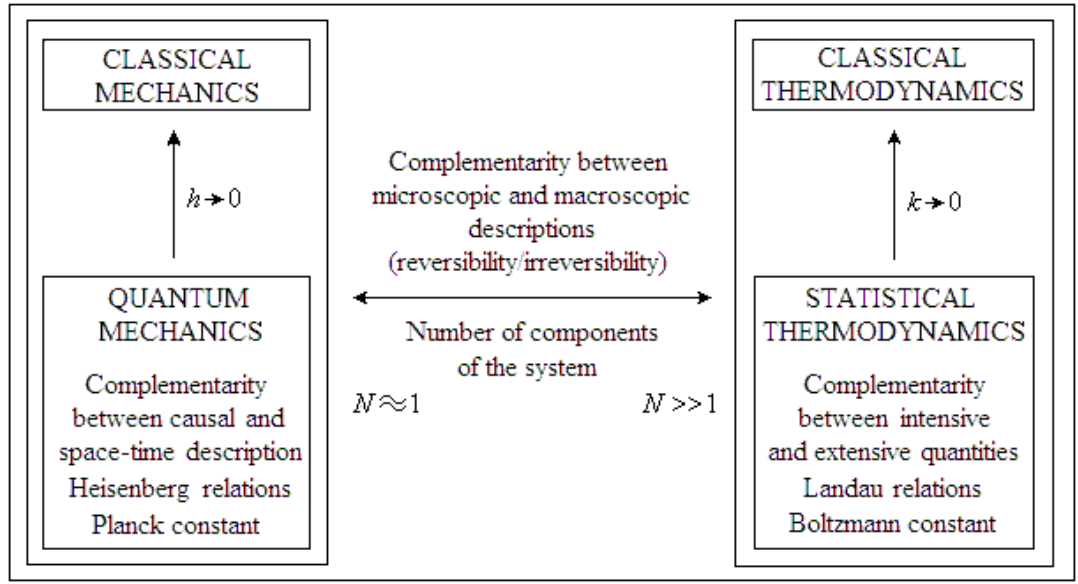

**Figure 3** – Complementarity and meta-complementarity

The first-order-relations, proper of QM and ST, refer to the microscopic and macroscopic levels, respectively and, in the limits $h\rightarrow 0$ and $k\rightarrow 0$, express the condition for the validity of the correspondent classical theories. The second-order complementarity, or meta-complementarity, clarifies the relationship between reversible-microscopic and irreversible-macroscopic descriptions and the impossibility to obtain a reduction of the latter to the former. There is not a hierarchical connection between dynamical and statistical descriptions: they have the same dignity and are alternatives to each other.

Both the first-order complementarity proper of the macroscopic level and the meta-complementarity have something to do with the request that $N$ must be very large. However, there is a difference between the two situations: the first one regards the dichotomy [*N large but finite – N infinite*], that differentiates the statistical and the "absolute" thermodynamic descriptions; the second refers to the alternative [*N very small – N very large*], that establishes the domains of predominance of the dynamical or the statistical description.

In conclusion, I would stress that, in order to accept the meta-theoretical framework presented here, it is not necessary to invoke two different ontologies for the micro- and macroscopic levels, nor to deny the unity of the physical world. We must only to take into account our position into a physical world that we can see in an objective manner only *from inside*, without the ability to grasp it with a single glance, as if we were gods. *Being* and *Becoming*, that Prigogine associates to the dynamical and thermodynamic descriptions, must be viewed as complementary modes of looking at nature.

**Acknowledgments**

I am grateful to prof. Enrico Giannetto for the support received during the preparation of this work at the Ph. D. School in Anthropology and Epistemology of Complexity of the Bergamo University.